\documentclass[11pt]{article}
\usepackage{coling2016}
\usepackage{times}
\usepackage{latexsym}

\usepackage{helvet}
\usepackage{courier}
\usepackage{graphicx}
\usepackage{algorithmic}
\usepackage{algorithm}
\usepackage{epsfig}
\usepackage{amssymb}
\usepackage{stfloats}
\usepackage{amsmath}
\usepackage{url}
\usepackage{multirow}
\usepackage{array}
\usepackage{verbatim}
\usepackage{subfigure}
\usepackage{epstopdf}
\usepackage{xcolor}

\newcommand{\ignore}[1]{{}}
\newcommand{\hankz}{\textcolor{black}}
\newcommand{\hjh}{\textcolor{black}}

\title{Distributed-Representation Based Hybrid Recommender System with Short Item Descriptions}

\author{Junhua He, Hankz Hankui Zhuo \and Jarvan Law  \\
Dept. of Computer Science, Sun Yat-Sen University, GuangZhou, China. 510006\\
hejunh@mail2.sysu.edu.cn, zhuohank@sysu.edu.cn,JarvanLaw@gmail.com}

\begin{document}

\maketitle

\begin{abstract}
Collaborative filtering (CF) aims to build a model from users' past behaviors and/or similar decisions made by other users, and use the model to recommend items for users. Despite of the success of previous collaborative filtering approaches, they are all based on the assumption that there are sufficient rating scores available for building high-quality recommendation models. In real world applications, however, it is often difficult to collect sufficient rating scores, especially when new items are introduced into the system, which makes the recommendation task challenging. We find that there are often ``short'' texts describing features of items, based on which we can approximate the similarity of items and make recommendation together with rating scores. In this paper we ``borrow'' the idea of vector representation of words to capture the information of short texts and embed it into a matrix factorization framework. We empirically show that our approach is effective by comparing it with state-of-the-art approaches.
\end{abstract}

\section{Introduction}

Recommender systems are a subclass of information filtering systems that seek to predict the rating or preference that a user would give to an item
\cite{DBLP:reference/rsh/2011}. Recommender systems have been applied to a variety of applications, e.g., movies, music, news, books, research articles, search queries, social tags, financial services \cite{DBLP:conf/aaai/FelfernigISZ07}, and Twitter followers \cite{conf/www/Gupta13}. In general there are three ways to design recommender systems \cite{DBLP:journals/tkde/AdomaviciusT05}, i.e., collaborative filtering \cite{DBLP:conf/uai/BreeseHK98}, content-based filtering \cite{DBLP:conf/nips/GopalanCB14}, and the hybrid filtering \cite{DBLP:journals/umuai/Burke02}. Our work follows the strand of hybrid filtering systems.

There have been works on hybrid filtering systems. For example, Saveski and Mantrach \cite{DBLP:conf/recsys/SaveskiM14} propose to exploit information from \emph{item document}, i.e., each item is assumed to be associated with a document, to help with recommendation based on the word frequency (or TF-IDF) in documents. Chen et al. present a topic-model based approach to utilize the context and item information \cite{DBLP:conf/aaai/ChenZWHL14} to help with recommendation. McAuley and Leskovec propose to build a hybrid recommender system by integrating information from review texts with rating scores \cite{DBLP:conf/recsys/McAuleyL13}. Despite the success of the previous approaches, they are based on the assumption that the text information is abundant enough for frequency mining or topic models extraction. When the item description is limited or short, e.g., only a few phrases or tags available, they will not work well since ``similar'' items with limited descriptions can be very different based on frequency. For example, an item described by ``a portable device'' should be similar to the item described by ``a light-weight and small equipment'', while they are very different based on frequency mining since they share very few words. There are indeed many applications, i.e., MovieLens\footnote{https://movielens.org} as shown in Table \ref{example-item}, where item descriptions are often short.
\begin{table}[!ht]
\caption{Examples of item descriptions in MovieLens}\label{example-item}
\centering
\begin{tabular}{|p{0.147\textwidth}|l|}
\hline
Item & Description \\ \hline\hline
Toy Story  & animation, children's, comedy \\ \hline
Jumanji  & adventure, children's, fantasy \\ \hline
Heat  & action, crime, thriller \\ \hline
Sabrina  & comedy, romance \\ \hline
Tom and Huck  & adventure, children's \\ \hline
Sudden Death  & action \\ \hline
GoldenEye  & action, adventure, thriller \\ \hline
\end{tabular}
\end{table}

In this paper, we aim to explore the similarity between short item descriptions by looking into the semantic relations between descriptions. Inspired by the vector representations of words \cite{word2vec}, which has been shown to be effective in capturing the semantic relations among words, we borrow the idea of vector representations to take advantage of short item descriptions to assist recommendation. We first build a matrix based on the vector representations of words, and then integrate the matrix into the rating scores to build a matrix factorization objective function. Finally we solve the objective function using an expectation-maximization algorithm to make item recommendation. We call our algorithm {\tt RECF}, which stands for hybrid \textbf{REC}ommender system based on collaborative \textbf{F}iltering with short reviews.

\ignore{such as linearly combining both systems \cite{Claypool99combiningcontent-based}, adding content-based characteristics to collaborative models \cite{DBLP:conf/aaai/MelvilleMN02}, adding collaborative characteristics to content-based models \cite{Nicholas99combiningcontent}, or building a unifying recommendation model \cite{DBLP:conf/uai/PopesculUPL01}. }
 
\ignore{Sparsity in recommendation systems reveals the fact that the observed rating scores are too few to build high-quality recommender systems.
Although labeling data in the form of ``like'' or ``dislike'' could be helpful for complementing rating scores  \cite{csvd}.
However, the labeling data may be sparse as well.
In this paper, we exploit the texts for describing the features and functionalities of items to help in such situations.
For example, we can recommend ``Apple iPad'' to users who have scored ``Microsoft Surface'' highly,
because the texts describing them can potentially be used to establish their similarities.
Previous methods to achieve this tend to be ineffective when semantic information of the words is not captured,
since item similarities may only be established by the combinations of different words.
For example, a ``portable'' device may be similar to a device that is both ``light-weight'' and ``small''. }

\ignore{Several examples with text descriptions are also shown in Table \ref{example-item} which are from the MovieLens\footnote{http://grouplens.org/datasets/movielens/} dataset. The first column is item names (i.e., movie names) and the second is item description.
From Table \ref{example-item} we can see that a user giving the highest score to ``Sudden Death'' indicates he likes ``action'' movies, while a user giving the highest score to ``Sabrina'' suggests he likes ``comedy'' and ``romance'' movies. These information about users would be helpful for further recommending items for them, e.g., preferentially recommending ``action'' movies for the first user, and ``comedy'' movies for the second user.}

\ignore{Word vectors \cite{word2vec} is a class of methods that aim to represent words using high dimensional vectors,
which are useful for capturing the relationships between different words.
The combination of word vectors has been previously shown to be capable of capturing complex semantic information.
For example, the summation of the word vectors of ``Japan" and ``Tokyo'' has been shown to be very close to that of ``Beijing'' and ``China''.
Hence, word vectors can be naturally used to capture not only syntactic but also semantic similarities between the items.

In this paper, we aim to exploit distributed representations of description data to assist with recommendation,
together with rating scores and labeling data that may be available.
We propose a novel algorithm called {\tt RECF},
which stands for learning distributed \textbf{RE}presentations for heterogeneous \textbf{C}ollaborative \textbf{F}iltering with item descriptions.
{\tt RECF} learns distributed representations of item descriptions using neural networks \cite{word2vec,DBLP:conf/icml/LeM14}\ignore{\cite{DBLP:conf/nips/MnihH08,DBLP:conf/acl/TurianRB10,word2vec,DBLP:conf/icml/LeM14,DBLP:conf/aaai/Huang15}, where each word is represented by a vector, concatenated or averaged with other word vectors in the context. The resulting vector is used to predict other words in the context}. Specifically, {\tt RECF} first learns a vector representation for each word in the item text,
and then computes an average over all the word vectors in the item description.  
{\tt RECF} then uses the new vectors to form a new matrix and builds a hybrid model to estimate unknown rating scores by combining this matrix with the rating and labeling data.}
\ignore{
To sum up,  the contributions of this paper are as follows: 
\begin{itemize}
\item  To the best of our knowledge, {\tt RECF} is the first approach that explores the usage of distributed representations of item descriptions in the collaborative filtering system. 
\item We build a general hybrid collaborative filtering framework by integrating item description (or text information), rating scores and labelings, which is extendable to incorporate more information, such as user profiles, in the framework. 
\item We empirically exhibit the effectiveness of {\tt RECF} by comparing to state-of-the-art recommender systems.  
\end{itemize}
}
\ignore{{\tt RECF} integrates ratings, labelings and item descriptions to perform recommendations as shown in Figure \ref{bg}. By constructing vector representations with descriptions, {\tt RECF} solves several problems that existing methods have difficulties with.
First, as mentioned above, item descriptions are often too short to construct accurate word frequency (TFIDF) representations. {\tt RECF} will borrow the idea of distributed representations of words~\cite{word2vec}, which is detailed later, to solve this problem. Based on this idea, we can select a separate corpus to construct word vector representations. The benefit is that this corpus does not necessarily have to be completely content-relevant with our data set. Any standard corpus such as  news articles would suffice.
Second, since short descriptions are content-limited and can be very noisy, how can we ensure that they can be incorporated properly to produce a positive influence? {\tt RECF} uses a parameter to control this influence, which is detailed in Section \ref{cal-control-lambda}. This parameter is dynamically changed based on the influence of item descriptions. In cases certain item descriptions are uninformative, they will be ignored by {\tt RECF} in the final recommendations.

\begin{figure}[!ht]
  \centering
  \includegraphics[width=0.4\textwidth]{bg.eps}
  \caption {messages used in {\tt RECF}}
  \label{bg}
\end{figure}
}
\ignore{
The rest of the paper is organized as follows. We first review previous work related to our approach. After that we discuss our problem formulation and present our approach in detail. We then evaluate our approach in benchmark datasets by comparing it with state-of-the-art methods. Finally, we conclude the paper with future work.}

\section{Related Work}
Our work is related to distributed representations of words. \hjh{
In earlier work, many models have been proposed to learn a distributed representation of words. Collobert and Weston~\cite{DBLP:conf/icml/CollobertW08} propose a single convolutional neural network called SENNA, to output a host of language processing predictions. Mnih and Hinton~\cite{DBLP:conf/nips/MnihH08} propose a fast hierarchical language model called HLBL, based on Log-Bilinear in \cite{DBLP:conf/icml/MnihH07} along with a simple feature-based algorithm, which outperforms non-hierarchical neural models in their evaluations. Mikolov~\cite{mikolov2012statistical} proposes a new statistical language model, RNNLM, based on RNN in \cite{DBLP:conf/interspeech/MikolovKBCK10}. Huang et al.~\cite{DBLP:conf/acl/HuangSMN12} propose a new model which increases the global context-aware to enrich the semantic information of words.
}
And Mikolov et al.~\cite{DBLP:journals/corr/abs-1301-3781} proposed two new models, CBOW and Skip-gram. Both models use a simple neural network architecture that aims to predict the neighbors of a word.
CBOW predicts the current words based on the context and Skip-gram tries to maximize the classification accuracy of a word based on another word in the same sentence.
Mikolov et al.~\cite{DBLP:journals/corr/abs-1301-3781} also proposed a new tool for learning word vectors called word2vec~\cite{word2vec}.
To improve the accuracy of the word representation,
Then in the following year, focusing on this technology of distributed representations. Frome et al. used it to make the language model pre-training of a new deep visual-semantic embedding model, as it has been shown to efficiently learn semantically-meaningful floating point representations of terms from unannotated text \cite{DBLP:conf/nips/FromeCSBDRM13}.
Mikolov et al. \cite{DBLP:journals/corr/MikolovLS13} developed a method that can automate the process of generating and extending dictionaries and phrase tables.
Le and Mikolov \cite{DBLP:conf/icml/LeM14} proposed an unsupervised algorithm that learns fixed-length feature representations from variable-length texts. Qiu et al. \cite{DBLP:conf/emnlp/QiuZL15} explored distributed representations of words to detect analogies.
In this paper, we exploit the distributed representation approach to transform item descriptions to vectors, and assist recommendation based on these vectors.

\ignore{
Our work is also related to recommender system. Content-based filtering is one of the well-known approach in recommender system which aims to learn a classifier for recommendation according to the similarity of items \cite{DBLP:reference/ml/MelvilleS10,DBLP:conf/adaptive/PazzaniB07,DBLP:conf/recsys/CantadorBV10,DBLP:journals/mta/SoaresV15}. It is difficult to distinguish the messages of items also it can't discover new preference for users.
Collaborative filtering \cite{DBLP:journals/umuai/Burke02,DBLP:journals/tois/HerlockerKTR04,DBLP:conf/cscwd/ChangLCTC14,DBLP:conf/recsys/AdamopoulosT13} is one of the most well-known approaches in recommender systems which make recommendations according to the relations among the interests of all users.
\hjh{
In various recommender systems, there are common techniques that play important roles. Probabilistic latent semantic analysis (pLSA) provides a probabilistic approach for characterizing latent associations among objects\cite{DBLP:journals/itiis/LiLH15,DBLP:journals/aicom/KagieLW09}, and then, clustering is used to group similar users or items according to their features\cite{DBLP:conf/icml/LongZWY06,DBLP:conf/icml/LongZWY06}.
}
A few examples of such recommender systems include CMF \cite{DBLP:conf/kdd/2008}, LMF \cite{DBLP:conf/kdd/2011} and FM \cite{DBLP:conf/icdm/Rendle10}.
}
\ignore{
\subsection{Transfer learning}
Despite the success of previous approaches, they are all based on the assumption that the rating data (for training) is rich enough for building a collaborative filtering model. In real world applications, \hjh{however, there often exists the challenge that the rating data is sparse, and thus collaborative filtering such as approaches mentioned previously are unable to make high-quality recommendations~\cite{DBLP:journals/corr/abs-1301-7363,DBLP:conf/aaai/PanXLY10}. }

\hjh{
Transfer learning (aiming to use the knowledge from different information modalities to help learning in target tasks) is helpful for solving the sparsity problem. In \cite{DBLP:journals/tkde/PanY10}, Pan and Yang identified three categories of transfer learning including inductive transfer learning, transductive transfer learning and unsupervised transfer learning with four types of knowledge to transfer.
}
\hjh{
The first type is the knowledge of instances. After first being addressed with a boosting algorithm called TrAdaBoost~\cite{DBLP:conf/icml/DaiYXY07}, other methods are proposed later on for transferring this type of knowledge. Migratory-Logit proposed by Liao et al.~\cite{DBLP:conf/icml/LiaoXC05}, for example, completes the labels on data in a target domain with the help of the source domain data. And, Jiang and Zhai~\cite{DBLP:conf/acl/JiangZ07} integrate the auxiliary data in SVM, largely improving the accuracy in classification.
The second type is the knowledge of feature representation. Lee et al.~\cite{DBLP:conf/icml/LeeCVK07} present a convex optimization algorithm that transfers the meta-priors among different tasks. Jebara~\cite{DBLP:conf/icml/Jebara04} provides a method for common feature selection or kernek selection for SVMs. Blitzer et al.~\cite{DBLP:conf/emnlp/BlitzerMP06} propose a structural correspondence learning method to automatically induce correspondences among features from different domains.
The third type is parameters. Bonilla et al.~\cite{DBLP:conf/nips/BonillaCW07} utilize Gaussian Processes, proposing a model to learn a shared covariance function on input-dependent features and a ``free-for'' covariance matrix over tasks. And Gao et al.~\cite{DBLP:conf/kdd/GaoFJH08} propose a locally weighted ensemble framework to combine multiple models for transfer learning with the weights dynamically assigned.
The last type of the knowledge we can transfer is about relations in domains. Mihalkova et al.~\cite{DBLP:conf/aaai/MihalkovaHM07} propose to transfer relational knowledge with a complete MLN transfer system. Davids and Domingos~\cite{DBLP:conf/icml/DavisD09} discover structural regularities in the source domain in the form of Markov logic formulas.
}

With its advantage, transfer learning has a great impact in recent studies. The researchers use all kinds of messages to transfer knowledge. For examples,
In recent studies, transfer learning is very popular and has a great impact. For example, Koenigstein et al.\cite{DBLP:conf/recsys/KoenigsteinDK11} used information about music tracks, albums, artists and genres to help with recommendation.
Moshfeghi et al. \cite{DBLP:conf/sigir/MoshfeghiPJ11} made use of item summary and reviews. 
Li et al. \cite{DBLP:conf/cikm/LiHZC10} used user query history, and combined it with purchasing and browsing activities. 
Gantner et al. \cite{DBLP:conf/icdm/GantnerDFRS10} combined information of both users and items to deal with the cold-start scenarios.
Pan and Yang \cite{csvd} improved the performance of recommendation by mapping auxiliary data to target data.

However, most of these approaches do not consider the semantic similarities between the combinations of different words.
Differing from these previous approaches, we exploit item descriptions by making use of distributed representations of words \cite{word2vec}
to assist with recommendation when rating and labeling data are sparse.
}

\section{Problem Formulation}
A \emph{rating} matrix is denoted by $R\in \{1,2,3,4,5,?\}^{N\times M}$, where $R_{uv}$ is a rating score given by user $u$ for item $v$, $N$ is the number of users, $M$ is the number of items, and the symbol ``?'' indicates no score is given by user $u$. An \emph{labeling} matrix is denoted by $L\in\{0,1,?\}^{N\times M}$, where $L_{uv}$ is the label given by user $u$ for item $v$, with the meaning of ``dislike'', ``like'' and ``unknown label'' for ``0'', ``1'' and ``?'', respectively. An \emph{item description} vector is denoted by $Q$, where $Q_v$ is composed of a set of words describing the properties of item $v$. Note that $Q_v$ can be an empty set $\emptyset$ suggesting no item description given to item $v$.
\ignore{
\begin{figure}[!ht]
  \centering
  \includegraphics[width=0.4\textwidth]{bg.eps}
  \caption {messages used in {\tt RECF}}
  \label{bg}
\end{figure}
}

Our recommender system can be defined by: given as input a rating matrix $R$, a labeling matrix $L$ and an item description vector $Q$, it aims to estimate \emph{unknown} rating scores ``?'' in $R$.
\ignore{
We use $Q$ to denote the item descriptions. $Q_k$, representing the $k^{th}$ line of the data, is composed of words for describing the $k^{th}$ item.
$C$ is the item description matrix transformed from $Q$, where $C_{v\cdot}$ represents the vector-form description of the $v^{th}$ item.
We denote the rating matrix as $R$,
where $R_{uv}$ is the rating value of the $u^{th}$ row (i.e., the $u^{th}$ user) and the $v^{th}$ column (i.e., $v{th}$ item) in $R$.
Similarly, $L$ is the labeling matrix, where $L_{uv}$ is the labeling value ``like" as $1$ and ``dislike" as $0$) of the $u^{th}$ row and the $v^{th}$ column in $L$.

Note that the heterogeneous data $R$, $L$, $C$ are the inputs of our learning algorithm, and we aim to learn the user latent features $U$, item latent features $V$ and the data-dependent effect $B_R$. Finally, with the output, we can make item recommendations based on $UB_RV^{T}$.}

\section{Our {\tt RECF} Algorithm} \label{algorithm}
In this section, we present our {\tt RECF} algorithm in detail. We first build distributed representations of item descriptions, and then integrate the ratings, labelings and distributed representations of item descriptions to build a bayesian model and learn parameters of the model to build the recommender system. An overview of {\tt RECF} is shown in Algorithm \ref{algorithm:overview}. We will address each step of Algorithm \ref{algorithm:overview} in detail in the subsequent sections.
\begin{algorithm}
\caption{The framework of our {\tt RECF} algorithm}\label{algorithm:overview}
\textbf{input:} ratings $R$, labelings $L$, item descriptions $Q$ \\
\textbf{output:} estimated ratings $\hat{R}$

\begin{algorithmic}[0]
\STATE 1: build representations $C$ from descriptions $Q$
\STATE 2: build hybrid model $\mathcal{M}$ based on $R$, $L$ and $C$
\STATE 3: learn the parameters of $\mathcal{M}$ with EM approach
\STATE \quad 3.0: initiate $U$, $V$, $B_R$, $B_L$ and $W_C$
\WHILE{the maximal iteration is not reached}
 \STATE 3.1: update $V$ using $U$, $B_R$, $B_L$ and $W_C$
 \STATE 3.2: update $U$ using $V$, $B_R$, $B_L$ and $W_C$
 \STATE 3.3: calculate $B_R$, $B_L$ and $W_C$ using $V$ and $U$
\ENDWHILE
\STATE 4: compute $\hat{R}=U^TB_RV$
\RETURN $\hat{R}$
\end{algorithmic}
\end{algorithm}

\subsection{Distributed representations of descriptions}
As the first step of Algorithm \ref{algorithm:overview}, we aim to build the distributed representations of item descriptions with $Q$ as input. 
We first learn the vector representations for words using the Skip-gram model with hierarchical softmax, which has been shown an efficient method for learning high-quality vector representations of words from unstructured corpora \cite{word2vec}. The objective of the Skip-gram model is to learn vector representations for predicting the surrounding words in a sentence or document. Given a corpus $\mathcal{C}$, composed of a sequence of training words $\langle w_1,w_2,\ldots,w_T\rangle$, where $T=|\mathcal{C}|$, the Skip-gram model maximizes the average log probability
\begin{equation}\label{skip-gram}
\frac{1}{T}\sum_{t=1}^T\sum_{-c\leq j\leq c,j\neq0}\log p(w_{t+j}|w_t),
\end{equation}
where $c$ is the size of the training window or context.

The basic probability $p(w_{t+j}|w_t)$ is defined by the hierarchical softmax, which uses a binary tree representation of the output layer with the $K$ words as its leaves and for each node, explicitly represents the relative probabilities of its child nodes \cite{word2vec}. For each leaf node, there is an unique path from the root to the node, and this path is used to estimate the probability of the word represented by the leaf node. There are no explicit output vector representations for words. Instead, each inner node has an output vector $v'_{n(w,j)}$, and the probability of a word being the output word is defined by
$p(w_{t+j}|w_t)=\prod_{i=1}^{L(w_{t+j})-1}\Big\{\sigma(\mathbb{I}(n(w_{t+j},i+1)= child(n(w_{t+j},i)))\cdot v_{n(w_{t+j},i)}\cdot v_{w_t})\Big\},
$
where $\sigma(x)=1/(1+\exp(-x))$. $L(w)$ is the length from the root to the word $w$ in the binary tree, e.g., $L(w)=4$ if there are four nodes from the root to $w$. $n(w,i)$ is the $i$th node from the root to $w$, e.g., $n(w,1)=root$ and $n(w,L(w))=w$. $child(n)$ is a fixed child (e.g., left child) of node $n$.  $v_{n}$ is the vector representation of the inner node $n$. $v_{w_t}$ is the input vector representation of word $w_t$. The identity function $\mathbb{I}(x)$ is 1 if $x$ is true; otherwise it is -1. We can thus build vector representations of words $w$, denoted by $vec(w)$, by maximizing Equation (\ref{skip-gram}) with corpora.

With vector representations of words, we calculate the overall representations of item descriptions by ``summarizing'' all words in each item description.
\ignore{ For more effectively using this information,
we transform it to vector form $C$ as the input of learning algorithm.
We use word2vec \cite{word2vec} as the tool for learning word vector representations.

First,  we select an English corpus and a Chinese corpus to train word vectors\footnote{The code for word2vec \cite{word2vec} is available at \url{https://code.google.com/p/word2vec/}.}.
During training, we set the dimension of word vectors as $10$ and other parameters as the default values.
Then, we replace all the item descriptions with their vector representations and compute a feature vector for each item.}There could be many different ways to ``summarize'' all words. In this paper we consider a straightforward way of computing the overall representations of item descriptions, i.e., calculating an average representation over all words in each item description. We have 
$C_v=\frac{1}{|Q_v|}\sum_{w\in Q_v}vec(w),$
where $Q_v$ is the set of words describing the properties of item $v$. If $Q_v=\emptyset$, $C_v$ is assigned with symbol ``?'' with the same meaning in $R$. Note that we assume the importance of different words in $Q_v$ is identical in describing item $v$. It is possible to extend it to considering different importance of words by introducing weights to words when the prior knowledge is provided. We call the resulting matrix $C=[C_1,C_2,\ldots,C_M]^T$ \emph{description matrix}.
\ignore{
\begin{enumerate}
\item \textbf{Average representation:}  We compute an average representation over all words in each item description, i.e.,
\[C_v=\frac{1}{|Q_v|}\sum_{w\in Q_v}vec(w),\]
where $Q_v$ is the set of words describing the properties of item $v$. If $Q_v=\emptyset$, $C_v$ is assigned with symbol ``?'' with the same meaning in $R$. Note that we assume the importance of different words in $Q_v$ is identical in describing item $v$. It is possible to extend it to considering different importance of words by introducing weights to words when the prior knowledge is provided.
\item \textbf{Central vector:} Central vector is the vector with minimum sum of distances from all other vectors.
We compute the central vector based on the following optimization function:
\begin{eqnarray}\label{toVec2}
\arg\min_{cv}\sum_{w\in Q_v}dist(cv, vec(w)),
\end{eqnarray}
where $cv$ is the central vector to be calculated, and $dist(cv, vec(w))$ is the Euclidean distance between $cv$ and $vec(w)$.
\end{enumerate}

We call the resulting matrix $C=[C_1,C_2,\ldots,C_M]^T$ \emph{description matrix}. We will exploit the first way as the default method of calculating the item description.We will exhibit the advantage and disadvantage of these two ways in the experiment.
}
\subsection{The hybrid model with item descriptions}
In Step 2 of Algorithm \ref{algorithm:overview}, we aim to build a hybrid model $\mathcal{M}$ to capture the underlying relations among ratings $R$, labelings $L$, and item descriptions $C$. 
\begin{figure}[!ht]
  \centering
  \includegraphics[width=0.35\textwidth]{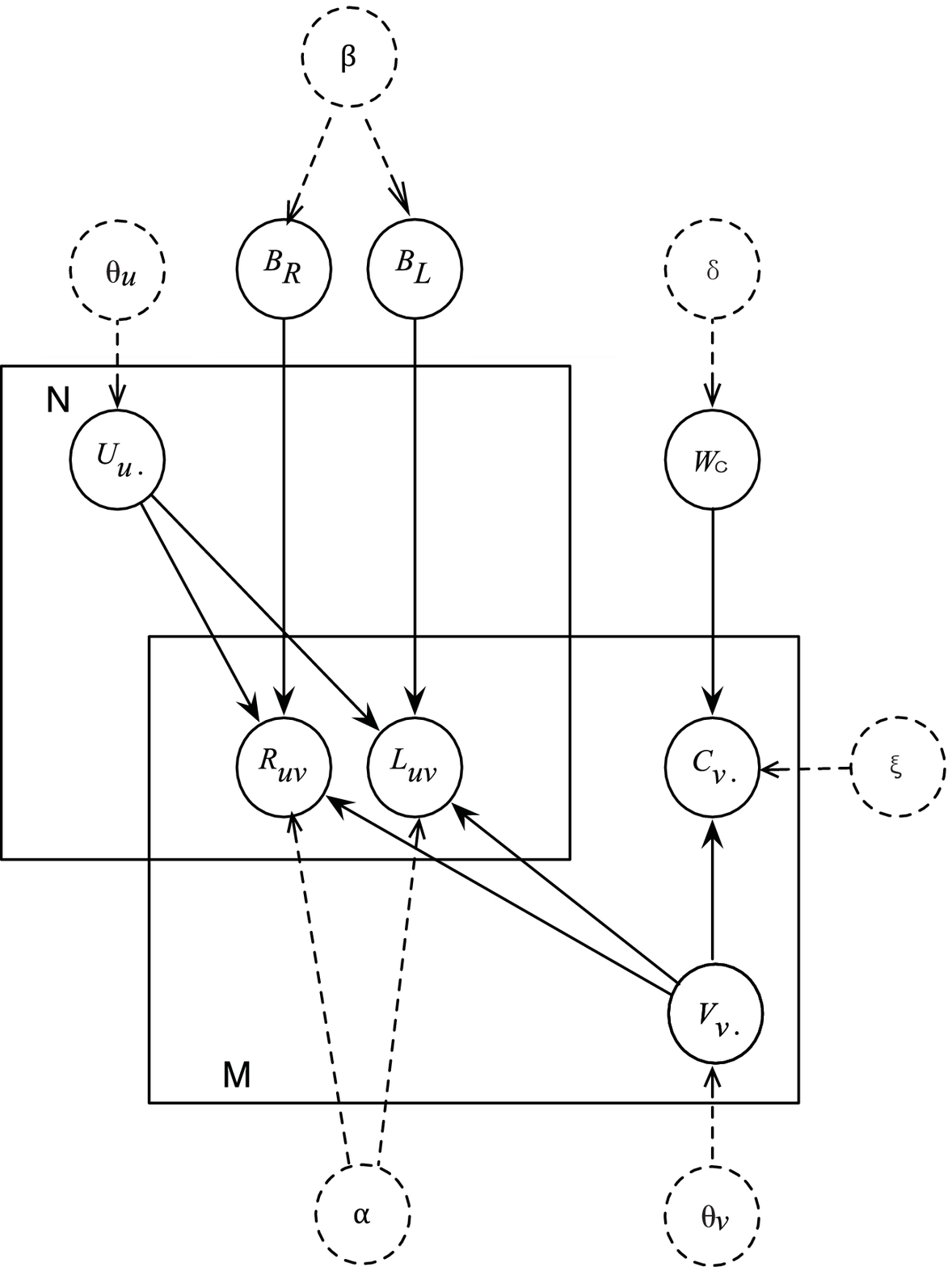}
  \caption{The hybrid model with ratings, labelings and item descriptions.}
  \label{framework}
\end{figure}
The framework of the hybrid model is shown in Figure \ref{framework}. 
The rationale of the hybrid model is based on the following four assumptions.\\
\textbf{Assumption 1:} Each user $u$ and item $v$ are characterized by an \emph{unknown} feature vector $U_u$ controlled by parameter $\theta_u$ and $V_v$ controlled by parameter $\theta_v$, respectively. The rating $R_{uv}$, which is controlled by parameter $\alpha$, is assumed to be resulted from bridging $U_u$ and $V_v$ with \emph{unknown} matrix $B_R$ controlled by parameter $\beta$. In other words, rating $R_{uv}$ should be close to $U_u B_R V_v^T$, i.e., $R_{uv}\sim U_u B_R V_v^T$. The similar idea is exploited by \cite{csvd}. This idea can be formulated by maximizing the conditional distribution below, assuming it follows a Gaussian distribution:
$p(R_{uv}|U_u,B_R,V_v,\alpha)=\mathcal{N}(R_{uv}|U_uB_RV_v^T,\alpha^{-1}\textbf{I})$
where
$\mathcal{N}(x|\mu,\alpha^{-1}\textbf{I})=\sqrt{\frac{\alpha}{2\pi}}\exp(-\alpha(x-\mu)^2).$
\\
\textbf{Assumption 2:} Likewise, the labeling $L_{uv}$, which is controlled by parameter $\alpha$, is assumed to be resulted from bridging $U_u$ and $V_v$ with \emph{unknown} matrix $B_L$ controlled by the same parameter $\beta$ of $B_R$, i.e., $L_{uv}\sim U_uB_L V_v^T$. We thus have
$p(L_{uv}|U_u,B_L,V_v,\alpha)=\mathcal{N}(L_{uv}|U_uB_LV_v^T,\alpha^{-1}\textbf{I}).$ 
\\
\textbf{Assumption 3:} The item description $C_v$, which is controlled by parameter $\xi$, is assumed to be resulted from the item features $V_v$ and \emph{unknown} matrix $W_C$ controlled by parameter $\delta$, i.e., $C_v\sim V_v^TW_C$. We thus have
$p(C_v|V_v,W_C,\xi)=\mathcal{N}(C_v|V_v^TW_C,\xi^{-1}\textbf{I}).$ 
\\
\textbf{Assumption 4:} Furthermore, we assume the distributions of $U_u$, $V_v$, $B_R$, $B_L$ and $W_C$ are
$p(U_u|\theta_u)=\mathcal{N}(U_u|\textbf{0},\theta_u^{-1}\textbf{I}),$
$p(V_v|\theta_v)=\mathcal{N}(V_v|\textbf{0},\theta_v^{-1}\textbf{I}),$
$p(B_R|\beta)=\mathcal{N}(B_R|\textbf{0},(\beta/q_R)^{-1}\textbf{I}),$
$p(B_L|\beta)=\mathcal{N}(B_L|\textbf{0},(\beta/q_L)^{-1}\textbf{I}),$
$p(W_C|\delta)=\mathcal{N}(W_C|\textbf{0},(\delta/q_C)^{-1}\textbf{I}),$
where $q_R$, $q_L$, and $q_C$ are numbers of not ``?'' elements in $R$, $L$ and $C$, respectively.

Based on the hybrid model shown in Figure \ref{framework}, our objective is to maximize the function as below:
\begin{equation}\label{objective}
\max_{U,V,B_R,B_L,W_C,} \log\mathcal{F}_R + \lambda_L\log\mathcal{F}_L + \lambda_C\log\mathcal{F}_C
\end{equation}
where $\lambda_L>0$ and $\lambda_C>0$ are tradeoff parameters to balance the ratings, labelings and item descriptions. $U\in\mathbb{R}^{n\times d}$ and $V\in\mathbb{R}^{m\times d}$ satisfy $U^TU=I$ and $V^TV=I$, respectively.
$\mathcal{F}_R$, $\mathcal{F}_L$ and $\mathcal{F}_C$ are defined by
$\mathcal{F}_R = \prod_{u,v} \Big[p(R_{uv}|U_u,B_R,V_v,\alpha) p(U_u|\theta_U)  p(V_v|\theta_V) p(B_R|\beta)\Big]^{x_{uv}}, \notag
$
$\mathcal{F}_L = \prod_{u,v}\Big[p(L_{uv}|U_u,B_L,V_v,\alpha) p(U_u|\theta_U)  p(V_v|\theta_V)p(B_L|\beta)\Big]^{y_{uv}}, \notag
$
and
$\mathcal{F}_C = \prod_v \Big[p(C_v|W_C,V_v,\xi) p(V_v|\theta_V)p(W_C|\delta)\Big]^{z_v}, \notag
$
where $x_{uv}$, $y_{uv}$ and $z_v$ are indicator variables for $R_{uv}$, $L_{uv}$ and $C_v$, respectively. If $R_{uv} =$ ``?'' (or $L_{uv}=$``?'' or $C_v=\emptyset$), then $x_{uv} = 0$ (or $y_{uv}=0$ or $z_v=0$); otherwise $x_{uv} = 1$ (or $y_{uv}=1$ or $z_v=1$).
\ignore{
\begin{eqnarray}
\label{indicator}
x_{uv}=\begin{cases}1 &\mbox{if ${R}_{uv}$ exists}\\0 &\mbox{if ${R}_{uv}$ not exists}\end{cases}\nonumber
\end{eqnarray}
}

Specifically, based on the Gaussian distributions given above, the log-posterior function of the ratings is shown below:
\begin{eqnarray}\label{log-rating}
\log\mathcal{F}_R =  -\sum_{u,v}x_{uv}[\frac{\alpha}{2}(R_{uv}-U_uB_RV_v^T)^2  +\frac{\theta_U}{2}\|U_u\|^2 +\frac{\theta_V}{2}\|V_v\|^2+\frac{\beta}{2q}\|B_R\|_F^2+K_R],
\end{eqnarray}
where $K_R =ln\sqrt{\frac{\alpha}{2\pi}}+ln\sqrt{\frac{\theta_U}{2\pi}}+ln\sqrt{\frac{\theta_V}{2\pi}}+ln\sqrt{\frac{\beta}{2q_R\pi}}$ is a constant.
Likewise, we can compute the log-posterior functions of the labelings $\log\mathcal{F}_L$ and descriptions $\log\mathcal{F}_C$. We can see the objective function Equation (\ref{objective}) can be reduced to a polynomial function. We will solve the optimization problem using an EM-style algorithm in the next subsection.
\subsection{The EM algorithm}
In Step 3 of Algorithm \ref{algorithm:overview} we aim to learn the parameters $B_R$, $B_L$, $W_C$, $U$ and $V$ using the EM approach. As the beginning of the EM approach, we initialize $U$ and $V$ using the SVD result of labelings $L$, since the labeling data $L$ describes users' ``high-level'' or general interest in items. After that we initialize $B_R$, $B_L$ and $W_C$ with Equations (\ref{estimateB}) and (\ref{estimateW}) using $U$ and $V$, which will be introduced in Section 4.3.2.

\ignore{Due to the sparsity of the data, the initializations of $U$ and $V$ from SVD results sometimes do not make sense. In such situations, we use $C$
to fill $V$ by setting
$V=CW_C^T(W_CW_C^T)^{-1}.$
}
\ignore{Moreover, we also initialize matrices of the indicator variables, denoted as $X$, $Y$ and $Z$, respectively, for $R$, $L$ and $C$ to indicate the location of the  missing data.}

\subsubsection{Learning $V$ and $U$} 
In Steps 3.1 and 3.2 of Algorithm \ref{algorithm:overview}, we aim to learn $V$ and $U$. Given $U$ and $B_R$, $B_L$,  $W_C$, we can update $V$ using gradient descent approach. We first simplify the optimization function from Equation (\ref{objective}), as shown below:
\begin{eqnarray}\label{optimizeV}
&&\!\!\!\!\!\!\!\!\!\!\! \min_{U,V}f = \min_{U,V}\frac{1}{2}\|X\odot(R-UB_RV^T)\|_F^2+\frac{\lambda_R}{2}\|Y\odot (L-UB_LV^T)\|_F^2+\frac{\lambda_C}{2}\|Z\odot(C-VW_C)\|_F^2\nonumber\\
&&\!\!\!\!\!\!\!\!\!\!\!  s.t.~~~~ U^TU=I,V^TV=I, 
\end{eqnarray} where $X=[x_{uv}]$, $Y=[y_{uv}]$, $Z=[z_v]$.
We then iteratively update $V$ and $U$ by 
$V = V - \gamma_1\frac{\partial{f}}{\partial{V}}$ and
$U = U - \gamma_2\frac{\partial{f}}{\partial{U}}$, 
where $\gamma_1$ and $\gamma_2$ are two learning constants.
\ignore{
\begin{eqnarray}
\nabla V =\frac{\partial{f}}{\partial{V}}=[X\odot(VB_R^TU^T-R^T)UB_R]+ \lambda_L[Y\odot \nonumber\\
\!\!\!\!\!\!(VB_L^TU^T-L^T)VB_L] + \lambda_C[Z\odot(VW_C-C)W_C^T], \nonumber
\end{eqnarray}
where the parameter $\gamma_V$ will be introduced in Section \ref{cal-gamma}.

\subsubsection{Learning $U$} Similarly, given $B_R$, $B_L$, $U$ and updated $V$, we can update $U$ from Equation (\ref{objective}) using the similar method with learning $V$. The final update function is given as follows:
\begin{eqnarray}\label{updateU}
U = U - \gamma_U\nabla U
\end{eqnarray}
where
\begin{eqnarray}
\nabla U=[X\odot(UB_RV^T-R)]VB_R^T+ \nonumber \\ \lambda_L[Y\odot(UB_LV^T-L)]VB_L^T, \nonumber
\end{eqnarray}
where the parameter $\gamma_U$ will be introduced in the next subsection.
\ignore{
\begin{eqnarray}
\gamma_U = \frac{-tr(t_{R1}^Tt_{R2})-\lambda_Ltr(t_{L1}^Tt_{L2})}{tr(t_{R2}^Tt_{R2})+\lambda_Ltr(t_{L2}^Tt_{L2})} \nonumber
\end{eqnarray}
where $t_{R1}=X\odot(R-UB_RV^T)$, $t_{L1}=Y\odot(L-UB_LV^T)$, $t_{R2}=X\odot(\nabla UB_RV^T)$ and $t_{L2}=Y\odot(\nabla UB_LV^T)$.}
}

\ignore{ 
\subsubsection{Calculating $\gamma_V$, $\gamma_U$} \label{cal-gamma}
As for $\gamma_V$, plugging the update rule in Equation (\ref{updateV}) into the objective function in Equation (\ref{optimizeV}), we have
\begin{eqnarray}
\!\!\!\!\!\!\!g(\gamma_V)=\frac{1}{2}\|X\odot(R-UB_RV^T)+\gamma_V(UB_R\nabla V^T)\|_F^2 \nonumber \\
+ \frac{\lambda_L}{2}\|Y\odot(L-UB_LV^T)+\gamma_V(UB_L\nabla V^T)\|_F^2 \nonumber \\
+ \frac{\lambda_C}{2}\|Z\odot(C-VW_C)+\gamma_V(\nabla VW_C)\|_F^2.~~~~~~~ \nonumber
\end{eqnarray}
Letting \[t_{R1}=X\odot(R-UB_RV^T),\]
\[t_{L1}=Y\odot(L-UB_LV^T),\]
\[t_{C1}=Z\odot(C-VW_C),\]
\[t_{R2}=X\odot(UB_R\nabla V^T),\]
\[t_{L2}=Y\odot(UB_L\nabla V^T),\]
\[t_{C2}=Z\odot(\nabla VW_C),\]
we have
\begin{eqnarray}
\frac{\partial{g(\gamma_V)}}{\partial{\gamma_V}}&=&tr(t_{R1}^Tt_{R2}) + \gamma_V tr(t_{R2}^Tt_{R2}) \nonumber\\
&+&\lambda_L[tr(t_{L1}^Tt_{L2}) + \gamma_V tr(t_{L2}^Tt_{L2})] \nonumber\\ 
&+&\lambda_C[tr(t_{C1}^Tt_{C2}) + \gamma_V tr(t_{C2}^Tt_{C2})]. \nonumber
\end{eqnarray}
By setting $\frac{\partial{g(\gamma_V)}}{\partial{\gamma_V}}=0$, we have
\begin{eqnarray}
\!\!\!\!\!\!\!\!\!\!\!\! \gamma_V =\frac{-tr(t_{R1}^Tt_{R2})-\lambda_Ltr(t_{L1}^Tt_{L2})-\lambda_Ctr(t_{C1}^Tt_{C2})}{tr(t_{R2}^Tt_{R2})+ \lambda_Ltr(t_{L2}^Tt_{L2})+ \lambda_Ctr(t_{C2}^Tt_{C2})}.
\end{eqnarray}
Similarly, we can compute $\gamma_U$ by,
\begin{eqnarray}
\gamma_U = \frac{-tr(t_{R1}^Tt_{R2})-\lambda_Ltr(t_{L1}^Tt_{L2})}{tr(t_{R2}^Tt_{R2})+\lambda_Ltr(t_{L2}^Tt_{L2})}.
\end{eqnarray}
}

\subsubsection{Calculating $B_R$, $B_L$ and $W_C$} 
In Step 3.3 of Algorithm \ref{algorithm:overview}, we compute $B_R$, $B_L$ and $W_C$ using $U$ and $V$. For $B_R$, we have the optimal function shown below,
$\min_{B_R}\frac{1}{2}\|X\odot(R-UB_RV^T)\|_F^2+\frac{\beta}{2}\|B_R\|_F^2. \nonumber
$
Letting $\mathbb{B}_R=vec(B_R)=[B_{R_{\cdot1}} \cdot\cdot\cdot B_{R_{\cdot d}}]$, $m_{ui}=vec(U_{u\cdot}^TV_{i\cdot})$, $\mathbb{R}=vec(R)$, where $vec(Y)$ indicates a vector built by concatenating columns of the matrix $Y$, we have the following equivalent problem,
$\min_{\mathbb{B}_R}\frac{1}{2}\|\mathbb{R}-\mathbb{M}\cdot \mathbb{B}_R\|_F^2+\frac{\beta}{2}\|\mathbb{B}_R\|_F^2, \nonumber
$
where $\mathbb{M}= [...m_{ui}...]^T$. Letting $\nabla \mathbb{B}_R=0$, we have
\begin{eqnarray}\label{estimateB}
vec(B_R) = \mathbb{B}_R =  (\mathbb{M}^T\mathbb{M}+\beta I)^{-1}\mathbb{M}^T\mathbb{R}.
\end{eqnarray}
Likewise, we have $vec(B_L)=(\mathbb{M}^T\mathbb{M}+\beta I)^{-1}\mathbb{M}^T\mathbb{L}$, where $\mathbb{L}=vec(L)$.
Finally, we can easily compute $B_R$ and $B_L$ from $vec(B_R)$ and $vec(B_L)$, respectively. 

Given $V$, we can estimate the parameter $W_C$ by optimizing the subject function from Equation (\ref{objective}). We have 
$\min_{W_C}\frac{\lambda_C}{2}\|Z\odot (C-VW_C)\|_F^2 + \frac{\delta}{2}\|W_C\|_F^2.
$
We calculate the gradient 
$\nabla {W_C} = -V^TC+V^TVW_C+\beta W_C$, and set $\nabla {W_C}=0$. As a result, we have 
\begin{equation}\label{estimateW}
W_C=(V^TV+\delta I)^{-1}V^TC.
\end{equation}

\subsubsection{Tradeoff between $\lambda_{L}$ and $\lambda_{C}$} \label{cal-control-lambda}
The initial values of the tradeoff parameters $\lambda_{L}$ and $\lambda_{C}$ are set before running the program, which are determined through repeated experiments. \hjh{During execution, $\lambda_{L}$ will remain the same while $C$ will change. The reason is that, the labeling data includes accurate information while item description matrix $C$ is obtained based on distributed representations of descriptions. When the labeling data is sparse, the noise issue with item descriptions may be worsen.} Thus, the positive influence of $C$ only plays in a macroscopic level but not in a microcosmic one. At the later period of convergence, continuing using $C$ may reduce the accuracy. In other words, the influence of $C$ should be gradually decreased as running the algorithm. We thus propose three options to adjust the value of $\lambda_C$, as shown below.
\begin{enumerate}
\item \textbf{Linear decline:} Linear decline is the simplest model to specify the declining, in which we compute $\lambda_C$ as follows:
\begin{eqnarray}
\!\!\!\!\!\!\!\!\lambda_C = \begin{cases}m-(iter-1)\cdot k &\mbox{if $iter < \frac{m}{k}+1$}\\0 &\mbox {else}\end{cases}
\end{eqnarray}
 where $m$ is the initial value of $\lambda_C$, $iter$ is the iteration and $k$ is the step size.
\item \textbf{Nonlinear decline:}
To emphasize the strong influence of $C$ in the early period, in nonlinear decline, the decreasing speed of $\lambda_C$  also decreases in the execution. We propose a simple model as follows:
\begin{eqnarray}
\label{nonlinear decline}
\lambda_C = m / iter
\end{eqnarray}
\item \textbf{Mutation:}
While the two methods mentioned above are easy to implement, the problem is that it is difficult to determine the step size.
Intuitively, if the number of iterations before convergence is large, we should adjust the value of $m$ to decrease the step size, thus extending the time of influence by $C$. 
Hence, we propose a method of mutating $C$ according to the convergence situation:
\begin{eqnarray}\label{cases}
\lambda_C = \begin{cases}m &\mbox{if before the first convergence}\\0 &\mbox {else}\end{cases}
\end{eqnarray}

The advantage of  this method is that we do not need to consider the convergence speed.
\end{enumerate}

\ignore{
Hence, We propose a method of mutating $C$ according to the convergence situation:
\begin{eqnarray}\label{cases}
\lambda_C = \begin{cases}m &\mbox{if before the first convergence}\\0 &\mbox {else}\end{cases}
\end{eqnarray}
where $m$ is the initial value.}

Finally, in Step 4 of Algorithm \ref{algorithm:overview}, we estimate values of ``?'' in  $R$ for recommendations by calculating $UB_RV^T$.

\section{Experiments}
In this section, we evaluate our {\tt RECF} algorithm using two datasets
MovieLens and Douban\footnote{\url{http://www.datatang.com/data/42832} and \url{http://www.datatang.com/data/44858}} by comparing it against other four algorithms, {\tt SVD} \cite{csvd}, {\tt CSVD} \cite{csvd}, {\tt CSVD+Binary} \cite{DBLP:conf/recsys/SaveskiM14} and {\tt CSVD+TFIDF} \cite{DBLP:conf/recsys/SaveskiM14}. {\tt SVD} is an approach that exploits just ratings for building models for item recommendations. {\tt CSVD} is an approach that exploits both ratings and labelings information for building models for item recommendations. {\tt CSVD+Binary} and {\tt CSVD+TFIDF} are two state-of-the-art approaches that exploit item description information for improving recommendation accuracy. They convert the item descriptions to \emph{Binary} matrix and tf-idf representations, respectively, and combine them with ratings together to build recommender systems \cite{DBLP:conf/recsys/SaveskiM14}. To make the comparison fair, we fed the labelings to the approaches by \cite{DBLP:conf/recsys/SaveskiM14}, resulting in {\tt CSVD+Binary} and {\tt CSVD+TFIDF}.
For both datasets MovieLens and Douban,
\ignore{there is a subset called ``ml-1m'' that contains 1000209 ratings from 0 to 5. There are 6040 users and 3952 movies in the subset. Each movie has its own description. The dataset used in the experiments is constructed as follows.
Among the 6040 users, we selected 3952 most frequent ones to use. Then, we filter 655982 ratings with values in [0,5], which are given by the 3952 users on 3952 movies.}
we randomly split the data into $n$ ($n$=3, 5, 10, 15, 20) subsets. We randomly selected one for training, one for building labeling data $L$ by setting $L_{uv}$ be $1$ if $R_{ui}>3$ and $L_{uv}$ be 0 otherwise (as done by \cite{csvd}). The other $n-2$ subsets are used for testing.
\ignore{The sparsity levels of each rating data and labeling data are 1.40\%, 0.84\%, 0.42\%, 0.28\% and 0.21\% respectively.}


We exploit two metrics to measure the performance, i.e., Mean Absolute Error (MAE) and Root Mean Square Error (RMSE), as shown below,
$MAE=\sum_{(u,i,r_{ui})\in T_E}|r_{ui}-\hat{r_{ui}}|/|T_E|$
and $RMSE=\sqrt{\sum_{(u,i,r_{ui})\in T_E}(r_{ui}-\hat{r}_{ui})^2/|T_E|},$
where $r_{ui}$ is the \emph{ground-truth} rating, $\hat{r}_{ui}$ is the predicted rating and $|T_E|$ is the number of testing ratings.

\hankz{
\begin{figure}[!ht]
\centering
  \includegraphics[width=0.48\textwidth]{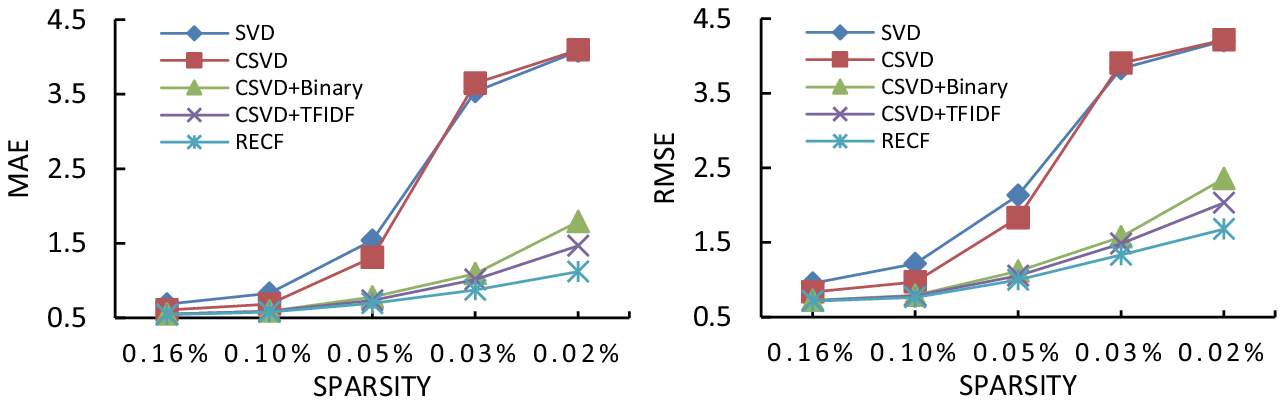}
  \caption{Performance of our {\tt RECF} algorithm and  w.r.t. sparsity of rating scores in dataset \textbf{Douban}.}
  \label{fig:Douban} 
\end{figure}
\begin{figure}[!ht]
\centering
  \includegraphics[width=0.49\textwidth]{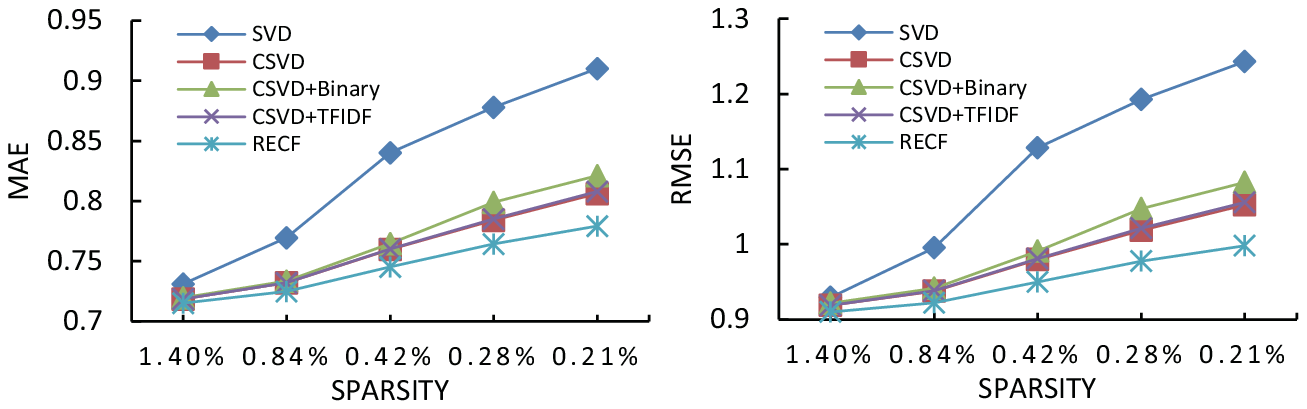}
  \caption{Performance of our {\tt RECF} algorithm and  w.r.t. sparsity of rating scores in dataset \textbf{MovieLens}.}
  \label{fig:MovieLen} 
\end{figure}
}
\subsection{Performance w.r.t. sparsity}
We first would like to see the performance with respect to different sparsities, by varying the percentage of available ratings (i.e., the rating scores given by users). We ran our {\tt RECF} algorithm and {\tt SVD}, {\tt CSVD}, {\tt CSVD+Binary}, {\tt CSVD+TFIDF} five times with different training and testing subsets and computed an average of accuracies. In our {\tt RECF} algorithm we set $\lambda_L$ to be 0.2 and $\lambda_C$ to be 2.5 in Equation (\ref{objective}). The results are shown in Figures \ref{fig:Douban} and \ref{fig:MovieLen}, \hankz{where we varied the sparsity from 1.4\% to 0.21\% in dataset \textbf{Douban}, and from 0.16\% to 0.02\% in dataset \textbf{MovieLens}, respectively. }

From the figures, we can see that both MAE and RMSE become larger when the percentage of ratings decreases in both datasets. This is consistent with our intuition since the fewer the ratings are, the larger the MAE and RMSE are.
Comparing different curves, {\tt CSVD+Binary}, {\tt CSVD+TFIDF} and {\tt RECF} algorithms generally perform better than {\tt SVD} and {\tt CSVD} in terms of MAE and RMSE, especially when the rating message is very sparse. This indicates item descriptions can indeed help improve the recommendation accuracy. However, in Douban field, we find that {\tt CSVD-TFIDF} performs almost the same as {\tt CSVD} while {\tt CSVD-Binary} even makes a negative effect on the result. The main reason is that the item descriptions we can use are only tags instead of long text descriptions. The item description information cannot be captured correctly by \emph{Binary} or \emph{tf-idf} matrix, which harms the recommendation accuracy. In contrast, our {\tt RECF} algorithm can better leverage these item description information based on distributed representations of words.

Furthermore, in both datasets, we can also observe that MAE (or RMSE) of {\tt SVD} (or {\tt CSVD}, {\tt CSVD+Binary}, {\tt CSVD+TFIDF}) increases faster than our {\tt RECF} algorithm as the percentage of rating scores decreases, i.e., the sparsity increases, which suggests that our {\tt RECF} algorithm functions even better, compared to the other four approaches, when the rating data is much sparser. This is because the impact of item descriptions relatively becomes larger when rating data decreases, resulting larger improvement of accuracies by item descriptions.
\ignore{In general, our {\tt RECF} algorithm performs better in terms of MAE and RMSE than both SVD and CSVD, which indicates item descriptions can indeed help improve the recommendation accuracy. We can also observe that MAE (or RMSE) of SVD (or CSVD) increases faster than our {\tt RECF} algorithm as the percentage of rating scores decreases, i.e., the sparsity increases, which suggests that {\tt RECF} functions even better, compared to SVD and CSVD, when the rating data is much sparser. This is because the impact of item descriptions relatively becomes larger when rating data decreases.} \ignore{However, with the change of the sparsity of training data, including rating and labeling data, the performance of {\tt RECF} and CSVD diverge. We can see that {\tt RECF} performs better than CSVD as the sparsity decreases.
When the sparsity is small, {\tt RECF} performs much better than CSVD,
which indicates that item descriptions can indeed help improve the accuracy when . }

\subsection{Tradeoff between $\lambda_C$ and $\lambda_L$}
\begin{figure}[!ht]
\centering
  \includegraphics[width=0.49\textwidth]{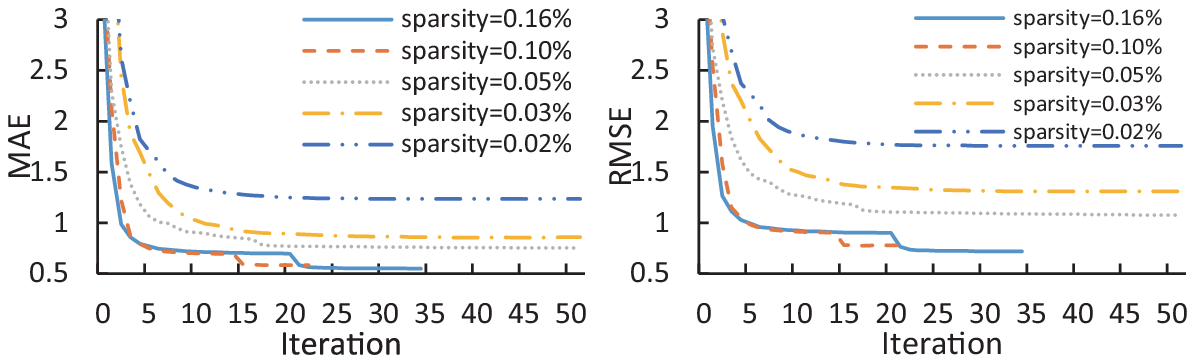}
  \caption{Impact of $\lambda_C$ and $\lambda_L$ in our {\tt RECF} algorithm in dataset \textbf{Douban}.}
  \label{fig:Douban_C} 
\end{figure}

\begin{figure}[!ht]
\centering
  \includegraphics[width=0.49\textwidth]{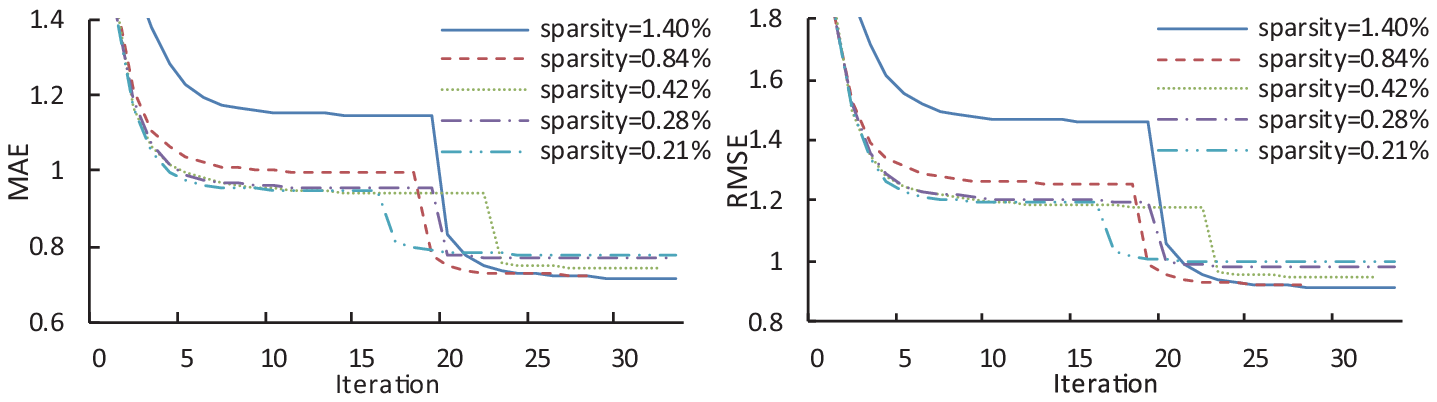}
  \caption{Impact of $\lambda_C$ and $\lambda_L$ in our {\tt RECF} algorithm in dataset \textbf{MovieLens}.}
  \label{fig:MovieLen_C} 
\end{figure}

Next, we would like to see the impact of $\lambda_C$ in Equation (\ref{objective}). We tuned the tradeoff between $\lambda_L$ and $\lambda_C$ by varying the value of $\lambda_C$ with respect to the number of iterations in {\tt RECF}, as presented by Equation (\ref{cases}). As we can see from Equation (\ref{cases}), the value of $\lambda_C$ is fixed to be $m$ before the first convergence and 0 once our {\tt RECF} algorithm converging, where $m$ is the preset initial value of $\lambda_C$. We fixed $m$ to be 2.5 and $\lambda_L$ to be 0.2 as done in the last subsection. We present the results in Figures \ref{fig:Douban_C} and \ref{fig:MovieLen_C}.

We find that the changes of performance (i.e., curves) can be divided into two stages, which indicate two phases of convergence. The first phase is for the tradeoff parameter of description matrix $C$, namely $\lambda_C$.
At the beginning of convergence, $C$ weights more than $L$ and dominates the convergence. In this period, the curve declines as expected.
However, as we can see from the figures, the curve may prematurely converge with a relatively low accuracy. The reason is that, due to the characteristics of $C$ -- capturing the similarity information in short descriptions with noise generated by word embedding, it may have a negative effect in a microcosmic level to get more accurate results.
\hankz{When we change $\lambda_C$ to $0$, i.e., $C$ no longer has any impact on the recommendation result, the curves go to another convergence stage, which verifies that $C$ mainly help improve the accuracy in the early stage by estimating values of ``?'' in $R$. Once the information from item descriptions $C$ has been encoded in $R$ and $L$ after the first convergence, the impact of item descriptions should be reduced (letting $\lambda_C$ be 0) and as a result, the impact of rating scores $R$ and labelings $L$ is relatively magnified to improve the recommendation accuracy. The rationale is that when the number of iterations reaches a threshold reducing the impact of description matrix $C$ could help avoiding overfitting when continuing running our {\tt RECF} algorithm. Note that setting $\lambda_C$ to be 0 indicates we do not need to update parameters $W_C$ in the objective function of Equation (\ref{objective}) and as a result the size of parameters to be learnt is reduced. In summary, $C$ should be weighed larger than $L$ in the early stage for quickly injecting it's impact on the learning process, and then reduced to zero to increase the impact of $R$ and $L$.}

\ignore{
\hankz{From the curves with different percentages of ratings less than 0.05\% in Figure \ref{fig:Douban_C}, there are no obvious second stage convergences when setting $\lambda_C$ to be zero. This is because the information available in $R$ is very little when the percentage is less than 0.05\%, and it does not have obvious help in improving the recommendation accuracy by setting $\lambda_C$ to be zero to increase the impact $R$ and $L$ after the first stage convergence. This is also verified by Figure \ref{fig:MovieLen_C} (in the dataset \emph{MovieLens}), where the second stage convergence is evident since all sparsities tested are larger than 0.05\%.}
\ignore{
\subsection{Different options of changing $\lambda_C$}
\hankz{
As mentioned in Subsection \ref{cal-control-lambda}, we proposed three options, i.e., \emph{Linear decline}, \emph{Nonlinear decline} and \emph{Mutation}, to calculate the value of $\lambda_C$. Different options may generate different results. We thus would like to conduct the evaluation of the three options by varying the number of iterations. We set the sparsity to be 0.42\% in MovieLens and 0.05\% in Douban, respectively. Likewise, we fixed $m$ to be 2.5 and $\lambda_L$ to be 0.2. The results are shown in Figures \ref{fig:BookDecline}.}
\begin{figure}[!htb]
  \centering
  \subfigure[iteration-MAE]{
    \label{fig:BookDeclineMAE} 
    \includegraphics[width=0.232\textwidth]{BookDeclineMAE_3.eps}}
  \subfigure[iteration-RMSE]{
    \label{fig:BookDeclineRMSE} 
    \includegraphics[width=0.232\textwidth]{BookDeclineRMSE_3.eps}}
  \caption{Three options of changing the impact of description matrix $C$ of our {\tt RECF} algorithm in the dataset \textbf{Douban}.}
  \label{fig:BookDecline} 
\end{figure}
}
\ignore{
\begin{figure}[!htb]
  \centering
  \subfigure[iteration-MAE]{
    \label{fig:MovieDeclineMAE} 
    \includegraphics[width=0.232\textwidth]{MovieDeclineMAE_3.eps}}
  \subfigure[iteration-RMSE]{
    \label{fig:MovieDeclineRMSE} 
    \includegraphics[width=0.232\textwidth]{MovieDeclineRMSE_3.eps}}
  \caption{Three options of changing the impact of description matrix $C$ of our {\tt RECF} algorithm w.r.t. the number of iterations in the dataset \textbf{MovieLens}.}
  \label{fig:MovieDecline} 
\end{figure}
}
\hankz{
From these results, in terms of accuracy, we can see that when the level of sparsity is still acceptable, these three strategies performed almost the same. And as expected, the curve of nonlinear decline drops the fastest, the curve of linear decline drops slower, and the curve from the mutational method can be seen as two phases. From this result, it seems that nonlinear decline is the best strategy.
However, when the rating and labeling data is very sparse (in Douban), nonlinear decline strategy resulted in a low performance. That is, $C$ plays a more important role in the message combination. However, the nonlinear decline function in Equation (\ref{nonlinear decline}) makes $\lambda_C$ decline so fast such that $C$ cannot introduce it effects. As a result, the program will overuse the messages of the sparse data $R$ and $L$, resulting in a lower performance.
For linear decline, this problem is less obvious. In terms of the convergence speed, we can see that linear decline converges much slower than the mutational method.
This analysis makes us eventually choose mutational decline in our algorithm.}
}
\ignore{
\begin{figure}[!htb]
  \centering
  \subfigure[$\lambda_C$-$MAE$]{
    \label{fig:BookLambdaMAE} 
    \includegraphics[width=0.232\textwidth]{BookLambdaMAE.eps}}
  \subfigure[$\lambda_C$-$RMSE$]{
    \label{fig:BookLambdaRMSE} 
    \includegraphics[width=0.232\textwidth]{BookLambdaRMSE.eps}}
  \caption{Performance of our {\tt RECF} algorithm w.r.t. different initial values of $\lambda_C$ in the dataset \textbf{Douban}, fixing $\lambda_L$ to be 0.2}
  \label{fig:Book_Lambda} 
\end{figure}

\begin{figure}[!htb]
  \centering
  \subfigure[$\lambda_C$-$MAE$]{
    \label{fig:MovieLambdaMAE} 
    \includegraphics[width=0.23\textwidth]{MovieLambdaMAE.eps}}
  \subfigure[$\lambda_C$-$RMSE$]{
    \label{fig:MovieLambdaRMSE} 
    \includegraphics[width=0.23\textwidth]{MovieLambdaRMSE.eps}}
  \caption{Performance of our {\tt RECF} algorithm w.r.t. different initial values of $\lambda_C$ in the dataset \textbf{MovieLens}, fixing $\lambda_L$ to be 0.2}
  \label{fig:Movie_Lambda} 
\end{figure}

\hankz{Furthermore, since the initial value of $\lambda_C$ is generally significant in influencing the performance of our {\tt RECF} algorithm, we conducted different options to choose a proper initial value. We fixed $\lambda_L$ to be 0.2, as done by the previous experiments, and varied the initial value of $\lambda_C$ (i.e., $m$ in Subsection \ref{cal-control-lambda}).  The results are shown in Figures \ref{fig:Book_Lambda} and \ref{fig:Movie_Lambda}. From the results, we can see that no matter what the sparsity of data was, the algorithm performed well when $\lambda_C$ was about 2.5. Some curves seem irregular such as the one when the sparsity is 0.02\% in Douban. However, in this situation, we can see that different values of $\lambda_C$ produced almost the same results, with differences less than 0.001. Clearly, this is caused by the level of sparsity in the training data. In such situations, varying $\lambda_C$ makes little difference.
Moreover, we can see that in these two different domains, the best-performing initial values of $\lambda_C$ are almost the same.
This may imply that we only need to fine-tune $\lambda_C$ in different domains.}
}

\ignore{
\begin{figure*}[!ht]
  \centering
  \subfigure[Sparsity-MAE]{
    \label{fig:BookMAE_new} 
    \includegraphics[width=0.48\textwidth]{BookMAE_new.eps}}
  \subfigure[Sparsity-RMSE]{
    \label{fig:BookRMSE_new} 
    \includegraphics[width=0.48\textwidth]{BookRMSE_new.eps}}
  \caption{Different ways of computing vector representations of item descriptions w.r.t. sparsity of rating scores in the dataset \textbf{Douban}}
  \label{fig:Book_new} 
\end{figure*}
\begin{figure*}[!ht]
  \centering
  \subfigure[Sparsity-MAE]{
    \label{fig:MovieMAE_new} 
    \includegraphics[width=0.48\textwidth]{MovieMAE_new.eps}}
  \subfigure[Sparsity-RMSE]{
    \label{fig:MovieRMSE_new} 
    \includegraphics[width=0.48\textwidth]{MovieRMSE_new.eps}}
  \caption{Different ways of computing vector representations of item descriptions w.r.t. sparsity of rating scores in the dataset \textbf{MovieLens}}
  \label{fig:Movie_new} 
\end{figure*}

\subsection{Feature vector of item description data $C$}
In this section we would like to test the accuracy with respect to different ways of calculating vector representations of item descriptions given vectors of words in each item description. As mentioned in Section \ref{cal-descrip-representation}, we explore two ways of calculating feature vector, i.e., \emph{average representation} and \emph{central vector}. We tested these two ways by varying sparsity of rating scores. Likewise, we fixed $\lambda_C$ and $\lambda_L$ to be 2.5 and 0.2, respectively. The results are shown in Figures \ref{fig:Book_new} and \ref{fig:Movie_new}, respectively.

From the figures, we observe that the two curves are almost coincide in the dataset \emph{MovieLen}. The reason is that the dataset \emph{MovieLen} is dense so that it weakens the impact of information from item description $C$. Besides, the number of words describing each item is limited, about three words per item description, which makes the feature vectors calculated by the two ways are similar. Likewise, in the dataset \emph{Douban}, the two ways of computing vector representations of item descriptions are almost identical. We thus randomly exploit one of the two ways in our {\tt RECF} algorithm (we exploit the first way in our {\tt RECF} algorithm, as mentioned in Subsection \ref{cal-descrip-representation}).
}
\ignore{
\subsection{Summary of the experiment}
Through analyzing the experimental results, we can make the following conclusions,
\begin{enumerate}
\item Item descriptions are very effective in solving the sparsity problem.
\item Distributed representations can indeed help improve recommendation accuracies by effectively transforming item descriptions to word vectors.
\item With the usage of item descriptions, our algorithm lead high performance in recommendation no matter how sparse the rating data and labeling are.
\end{enumerate}
}

\section{Conclusion}
In this paper, we propose a novel algorithm {\tt RECF} to explore item descriptions to help improve the recommendation accuracy using distributed representations of item descriptions.
Using this vector representation, we transform the item descriptions into vector representations, and combine them with rating and labeling data to build a hybrid recommender system. We exhibit that our {\tt RECF} approach is effective by comparing with the state-of-the-art approaches that exploit item descriptions. 
In the future, we would like to explore more information in our algorithm framework, such as user profiles or reviews, to further improve recommendation accuracies.

\newpage
\bibliographystyle{acl}
\bibliography{emnlp2016}

\end{document}